\documentclass[conference]{IEEEtran}
\IEEEoverridecommandlockouts
\usepackage[T1]{fontenc}
\usepackage[utf8]{inputenc}
\usepackage{cite}
\usepackage{graphicx}
\usepackage{booktabs}
\usepackage{tabularx}
\usepackage{threeparttable}
\usepackage{xcolor}
\usepackage{pifont}
\usepackage{amsmath}
\usepackage{array}
\usepackage{multirow}
\usepackage{enumitem}
\usepackage{amssymb}
\usepackage{url}
\usepackage[hidelinks]{hyperref}
\usepackage{marvosym}
\begin{document}
\title{BYOT-CPS: A Hybrid Cyber-Physical Systems Testbed for IoT Security Assessment and Platform Evaluation}
\author{
  \IEEEauthorblockN{
    Yan Lin Aung\textsuperscript{*,(\Letter)},
    Nelson Che Neba\textsuperscript{*}
  }
  \IEEEauthorblockA{
    University of Derby, United Kingdom \\
    y.aung@derby.ac.uk, n.cheneba1@unimail.derby.ac.uk
  }
}
%
%
\maketitle
\begingroup\renewcommand\thefootnote{*}
\footnotetext{Both authors contributed equally.}
\endgroup
\begin{abstract}
Internet of Things (IoT) security research continues to face a methodological gap between scalable virtual experimentation and realistic device behaviour.
While pure simulation and emulation platforms provide control, repeatability, and scale, they do not fully reproduce firmware-specific behaviours, hardware characteristics, and vendor implementation weaknesses that frequently determine real-world exploitability.
Conversely, physical-only testbeds provide realism but are costly to assemble, difficult to reconfigure, and hard to replicate across institutions.
This paper presents Build Your Own Cyber-Physical Systems Testbed (BYOT-CPS), a hybrid cyber-physical testbed that connects real IoT devices to virtualised network infrastructure built on GNS3.
BYOT-CPS is designed to support security experimentation, education, and independent evaluation of commercial IoT security platforms within a controlled environment that preserves authentic device behaviour.
Six requirements for such a testbed are defined: fidelity, heterogeneity, scalability, reproducibility, extensibility, and independence.
A prototype deployment integrating smart bulbs, smart plugs, switches, and IP cameras with virtual enterprise, server, attack, and monitoring zones is used to demonstrate hybrid connectivity, penetration testing workflows, a Mirai-style denial-of-service attack, traffic monitoring, and controlled device manipulation.
The evidence presented constitutes a feasibility validation of the framework rather than a large-scale comparative benchmark.
Within that scope, BYOT-CPS offers a practical middle ground between emulation-only research environments and costly physical laboratories while positioning vendor-neutral platform evaluation as a forward-looking design objective.
\end{abstract}
\section{Introduction}
The continued growth of the Internet of Things (IoT) has expanded the attack surface of modern organisations well beyond conventional endpoints and servers~\cite{iot-analytics-25,statista-25}.
Consumer, enterprise, and industrial deployments now encompass large numbers of connected cameras, lighting systems, plugs, sensors, controllers, and embedded appliances that commonly ship with weak default configurations, limited update support, and vendor-specific protocol stacks.
These characteristics have made IoT devices persistent targets for credential abuse, remote compromise, botnet recruitment, and lateral movement campaigns as demonstrated by Mirai and its successors~\cite{mirai-17}\cite{meneghello-19}.

Despite this threat landscape, rigorous IoT security experimentation remains methodologically constrained.
Existing research infrastructure tends to favour either scale or realism but rarely both.
Simulation and emulation frameworks such as Cooja/Contiki~\cite{cooja-06}, NS-3~\cite{ns3-10}, OMNeT++~\cite{omnet-08}, CORE~\cite{core-08}, and Mininet~\cite{mininet-10} support reproducible experimentation and large-scale topology construction, but they abstract away physical behaviours that most significantly affect security evaluation.
Physical testbeds, by contrast, expose genuine firmware flaws, management interface weaknesses, and protocol implementation anomalies, yet are costly to assemble, difficult to scale, and hard to reproduce across institutions.

A second challenge concerns independent evaluation.
Commercial IoT security platforms claim capabilities such as asset discovery, device classification, vulnerability detection, and behavioural monitoring; however organisations typically assess these claims through vendor-led demonstrations rather than controlled, reproducible experimentation.
The absence of a shared evaluation environment makes it difficult for researchers, educators, and prospective adopters to compare platforms under consistent conditions or to assess detection coverage against realistic attack scenarios.

This paper presents Build Your Own Cyber-Physical Systems Testbed (BYOT-CPS), a hybrid cyber-physical testbed designed to address these challenges.
BYOT-CPS combines virtual network infrastructure with real IoT hardware, enabling physical devices to participate as first-class nodes within an emulated topology.
The framework supports attack-and-defence experimentation, cybersecurity training, and the prospective independent evaluation of IoT security platforms without requiring deployment in live production environments.
Three linked objectives are addressed: defining a hybrid architecture for IoT security experimentation; showing, through an initial prototype, that such an architecture can support representative offensive and monitoring workflows against real devices; and treating vendor-neutral platform evaluation as a primary design objective rather than an ancillary consideration.

The paper makes the following contributions:
\begin{itemize}[leftmargin=*]
\item It defines a hybrid IoT security testbed architecture that integrates commercial-off-the-shelf (COTS) devices with virtualised network infrastructure while preserving realistic routing, segmentation, and monitoring behaviours.
\item It proposes six requirements for hybrid IoT security testbeds: fidelity, heterogeneity, scalability, reproducibility, extensibility, and independence, and uses them to position BYOT-CPS relative to existing approaches.
\item It presents a prototype implementation and initial feasibility experiments encompassing hybrid connectivity, vulnerability assessment, attack simulation, traffic monitoring, and controlled cyber-physical effects.
\end{itemize}

The core contribution of BYOT-CPS is the integration of real-device fidelity, flexible virtual topology construction, and explicit support for independent IoT security platform evaluation within a reproducible framework.

The remainder of the paper is organised as follows. Section~\ref{sec:requirements} defines design requirements; Section~\ref{sec:related} reviews related work; Section~\ref{sec:architecture} presents the BYOT-CPS architecture; Section~\ref{sec:evaluation} reports feasibility results.
Section~\ref{sec:discussion} discusses limitations and future work, and Section~\ref{sec:conclusion} concludes the paper.
\section{Hybrid IoT Security Testbed Requirements}\label{sec:requirements}
A hybrid IoT security testbed must satisfy requirements drawn from both network security experimentation and cyber-physical realism.
The literature consistently emphasises fidelity, heterogeneity, scalability, monitoring capability, and reproducibility~\cite{gotham-22,epic-13,attack-graphs-06,islam-15}.
Hybrid IoT environments introduce additional constraints as meaningful experimentation depends on interactions between virtual infrastructure and real devices rather than in isolation.
This paper defines six requirements to guide the design and evaluation of BYOT-CPS.
\subsection{Fidelity}
Fidelity concerns how closely the testbed reproduces the behaviour of the systems under study. 
In this context, fidelity encompasses three dimensions.
\emph{Node fidelity} requires the use of genuine operating systems, firmware, and services rather than behavioural abstractions.
\emph{Attacker fidelity} requires that experiments are conducted using real offensive tools, including scanners, exploitation frameworks, and, where appropriate, controlled malware or botnet components.
\emph{Topology fidelity} requires multi-segment network structures with configurable routing, switching, delay, loss, and monitoring properties.
This requirement is central to IoT security, as practically relevant weaknesses often arise from firmware behaviour, management interfaces, and protocol implementation details that are not preserved in purely simulated endpoints.
\subsection{Heterogeneity}
IoT ecosystems are inherently heterogeneous. 
Device vendors, communication protocols, management models, and security postures vary widely across deployments.
A useful testbed must therefore support mixed populations of devices and services, diverse traffic patterns, and multiple classes of attacks.
Heterogeneity is important not only for realism but also for the ecological validity of platform evaluation and the utility of any datasets generated within the environment.
\subsection{Scalability}
Scalability encompasses both infrastructural and organisational dimensions. 
At the infrastructure level, the framework must support growth from small laboratory deployments to larger multi-zone environments without requiring a fully physical build-out.
At the organisational level, it must not depend on a single proprietary or niche execution platform.
BYOT-CPS therefore treats portability across network emulation environments as an integral aspect of scalability rather than as a secondary implementation detail.
\subsection{Reproducibility}
Security research depends on reproducible experimental setups.
A hybrid testbed must permit topology capture, node configuration sharing, and scenario documentation sufficient for independent recreation by other researchers. 
Reproducibility is particularly significant in IoT research because without it the practical value of experimental results remains tied to a single local hardware inventory.
\subsection{Extensibility}
IoT security evolves rapidly, with new devices, attack techniques, and evaluation workflows emerging continuously.
A testbed architecture must support the addition of new physical devices, traffic generation tools, attack scenarios, and defensive platforms without requiring structural redesign.
Extensibility is important not only for long-term research value but also for educational use, where successive cohorts may require different scenarios and tooling.
\subsection{Independence}
The sixth requirement, and the one that most clearly distinguishes BYOT-CPS, is independence. 
A framework used to assess commercial security platforms must be vendor-neutral in its design, operation, and outputs.
It should enable competing tools to be evaluated under identical conditions and must not rely on proprietary orchestration logic controlled by a single vendor. 
This requirement addresses a practical gap in current IoT security practice: organisations need an impartial environment in which to assess platform claims prior to deployment.
\section{Related Work}\label{sec:related}
The literature relevant to BYOT-CPS spans network simulation, hybrid cyber ranges, IoT traffic generation, and commercial platform assessment.
While these areas provide useful components and precedents, no existing approach addresses all six requirements simultaneously as primary design objectives.
\subsection{Simulation and Emulation Frameworks}
General-purpose simulation and emulation platforms such as Cooja/Contiki-NG~\cite{cooja-06}, NS-3~\cite{ns3-10}, OMNeT++~\cite{omnet-08}, CORE~\cite{core-08}, and Mininet~\cite{mininet-10} remain foundational in networking and IoT research.
Their strengths are well established: repeatability, controllable topologies, and the ability to scale experiments beyond what is feasible with purely physical hardware.
However, these platforms do not natively reproduce the firmware, management interfaces, timing characteristics, or communication patterns of commercial IoT devices.
While this limitation may be acceptable for protocol research, it is often decisive for vulnerability assessment and realistic detection evaluation.

GNS3~\cite{gns3} and EVE-NG~\cite{eve-ng-25} occupy a more realistic position on the fidelity spectrum, supporting full network emulation, virtual appliances, and interaction with external interfaces.
They therefore provide strong foundations for hybrid experimentation.
However, they are infrastructure platforms rather than complete IoT security testbeds and do not, on their own, define a methodology for integrating physical IoT devices, structuring attack scenarios, or supporting vendor-neutral platform evaluation.

\subsection{IoT Security Testbeds and Hybrid Approaches}
Among prior research testbeds, the Gotham testbed~\cite{gotham-22} is the closest conceptual precedent to BYOT-CPS.
Gotham provides a scalable and reproducible IoT security environment with emulated devices, diverse protocols, and realistic offensive tooling.
Its contribution is substantial in terms of attacker fidelity and experimental repeatability.
However, it remains an emulation-centric environment and does not incorporate physical IoT hardware.
As a result, it cannot expose firmware-specific vulnerabilities or authentic cyber-physical device behaviour.
Balto et al. describe a hybrid IoT cyber range recognising the tension between physical fidelity and virtual scalability~\cite{hybrid-23}.
While this work supports the general case for hybrid environments, it does not demonstrate the same emphasis on multi-platform portability or the independent evaluation of commercial security platforms.
BYOT-CPS builds on the hybrid cyber range concept while establishing real-device participation and evaluation independence as explicit design objectives.
Container-based IoT environments have also appeared in the literature as lightweight and reproducible alternatives for physical deployments~\cite{container-19}.
Such approaches are useful for service emulation and traffic replay but remain approximations.
While containers can mimic application-layer services, they cannot replicate the exact firmware and hardware interactions of commercial devices.

\subsection{Traffic Generation and Commercial Evaluation}
Traffic generation tools such as MQTTset~\cite{mqttset-20} and GothX~\cite{gothx-24} support dataset production for intrusion detection and machine learning research.
They are valuable complements to a hybrid testbed, broadening behavioural coverage and enabling the generation of labelled traffic under controlled conditions.
However, they are not complete experimentation environments, as they do not provide topology control, real-device integration, or platform evaluation workflows.
Sandia’s Emulytics combines emulation, simulation, physical testbeds, and quantitative analysis across multiple cyber domains~\cite{emulytics}.
It targets large-scale experimentation, whereas BYOT-CPS emphasises real COTS IoT devices, practical attack and monitoring workflows, and vendor-neutral commercial platform evaluation within a common hybrid topology.
On the commercial side, IoT security platforms from vendors such as Armis, Claroty, Forescout, Cisco, Palo Alto Networks, and Nozomi Networks offer strong operational capabilities but limited transparency regarding independent validation.
Evaluations are typically conducted under vendor-controlled conditions.
BYOT-CPS addresses this practical gap by providing a controlled environment in which such platforms can be assessed against a common hybrid topology and device set, although systematic product benchmarking remains outside the scope of this paper.
\subsection{Comparative Positioning}
\begin{table*}[!t]
  \centering  
  \begin{threeparttable}
  \caption{Comparison of BYOT-CPS against related frameworks across the six requirements defined in Section~\ref{sec:requirements}.}
  \label{tab:comparison}
  \begin{tabular}{>  
  {\raggedright\arraybackslash}p{3.25cm}|>
  {\centering\arraybackslash}p{1.35cm}|>
  {\centering\arraybackslash}p{1.20cm}|>
  {\centering\arraybackslash}p{0.90cm}|>
  {\centering\arraybackslash}p{0.80cm}|>
  {\centering\arraybackslash}p{1.20cm}|>
  {\centering\arraybackslash}p{0.80cm}|>
  {\centering\arraybackslash}p{0.90cm}|>
  {\centering\arraybackslash}p{1.10cm}|>
  {\centering\arraybackslash}p{1.40cm}
  }
  \toprule
  \textbf{Property} & {NS-3 / OMNeT++} & \textbf{Cooja / Contiki} & \textbf{Gotham} & \textbf{GNS3} & \textbf{MQTTset} & \textbf{GothX} & Balto et al.~\cite{hybrid-23} & Emulytics & \textbf{BYOT-CPS} \\
  \midrule
  Real Device Behaviour\textsuperscript{\ding{172}}     & \ding{55}  & \ding{55}  & \ding{55} & \ding{109} & \ding{55}  & \ding{55}  & \ding{109} & \ding{109} & \ding{51} \\
  Attacker Tooling\textsuperscript{\ding{172}}          & \ding{55}  & \ding{55}  & \ding{51} & \ding{109} & \ding{55}  & \ding{55}  & \ding{109} & \ding{51}  & \ding{51} \\
  Complex Topology\textsuperscript{\ding{172}}          & \ding{51}  & \ding{55}  & \ding{51} & \ding{51}  & \ding{55}  & \ding{55}  & \ding{109} & \ding{51}  & \ding{51} \\
  Devices and Protocols\textsuperscript{\ding{173}}     & \ding{109} & \ding{55}  & \ding{51} & \ding{109} & \ding{109} & \ding{109} & \ding{109} & \ding{51}  & \ding{51} \\
  Attack Diversity\textsuperscript{\ding{173}}          & \ding{55}  & \ding{55}  & \ding{51} & \ding{109} & \ding{55}  & \ding{109} & \ding{109} & \ding{51}  & \ding{51} \\
  Topology Scalability\textsuperscript{\ding{174}}      & \ding{51}  & \ding{109} & \ding{51} & \ding{51}  & \ding{55}  & \ding{55}  & \ding{109} & \ding{51}  & \ding{51} \\
  Multi-Platform Support\textsuperscript{\ding{174}}    & \ding{55}  & \ding{55}  & \ding{55} & \ding{55}  & \ding{55}  & \ding{55}  & \ding{55}  & \ding{109} & \ding{51} \\
  Shareable\textsuperscript{\ding{175}}                 & \ding{51}  & \ding{51}  & \ding{51} & \ding{109} & \ding{51}  & \ding{109} & \ding{109} & \ding{109} & \ding{51} \\
  Third-Party Tool\textsuperscript{\ding{176}}          & \ding{109} & \ding{109} & \ding{51} & \ding{51}  & \ding{55}  & \ding{109} & \ding{109} & \ding{109} & \ding{51} \\
  Vendor-Neutral Evaluation\textsuperscript{\ding{177}} & \ding{55}  & \ding{55}  & \ding{55} & \ding{55}  & \ding{55}  & \ding{55}  & \ding{55}  & \ding{109} & \ding{51} \\
  \bottomrule
  \end{tabular}
  \begin{tablenotes}[para]
      \textsuperscript{\ding{172}}Fidelity, 
      \textsuperscript{\ding{173}}Heterogeneity, 
      \textsuperscript{\ding{174}}Scalability, 
      \textsuperscript{\ding{175}}Reproducibility,
      \textsuperscript{\ding{176}}Extensibility, 
      \textsuperscript{\ding{177}}Independence.

      \textsuperscript{\ding{55}}Not a stated focus, \textsuperscript{\ding{109}}Partially supported or indirectly supported, \textsuperscript{\ding{51}}Directly supported.
  \end{tablenotes}
  \end{threeparttable}
\end{table*}
Table~\ref{tab:comparison} provides a qualitative positioning of BYOT-CPS relative to representative prior work, based on capabilities explicitly described in the cited literature and public documentation.
The intent is not to diminish the value of simulation-only or emulation-centric environments, but to clarify that BYOT-CPS addresses a distinct problem: enabling security evaluation under conditions that preserve both device realism and flexible network composition.
%
%
\section{BYOT-CPS Architecture and Design}\label{sec:architecture}
BYOT-CPS is designed as a layered hybrid environment that integrates physical IoT devices with emulated network infrastructure while preserving the properties required for credible security experimentation.
The architecture is deliberately modular, enabling deployments to scale from a single-user academic laboratory to larger environments supporting defensive tooling and commercial platform evaluation.
\subsection{Design Principles}
Five principles guided the framework design, operationalising the requirements defined in Section~\ref{sec:requirements} at the architectural level.
\emph{Realism} requires the direct participation of COTS IoT hardware so that observed behaviour reflects actual firmware and device implementation.
\emph{Scalability} is achieved through virtual infrastructure hosted in GNS3.
\emph{Modularity} permits zones, services, and devices to be added or removed according to experimental objectives.
\emph{Isolation} ensures that offensive activity remains contained within the laboratory environment.
\emph{Accessibility} keeps the framework deployable on commodity hardware and standard tooling, reducing barriers to adoption in academic and resource-constrained settings.
\subsection{Layered System Model}
\begin{figure*}
  \centering
  \includegraphics[width=0.70\textwidth]{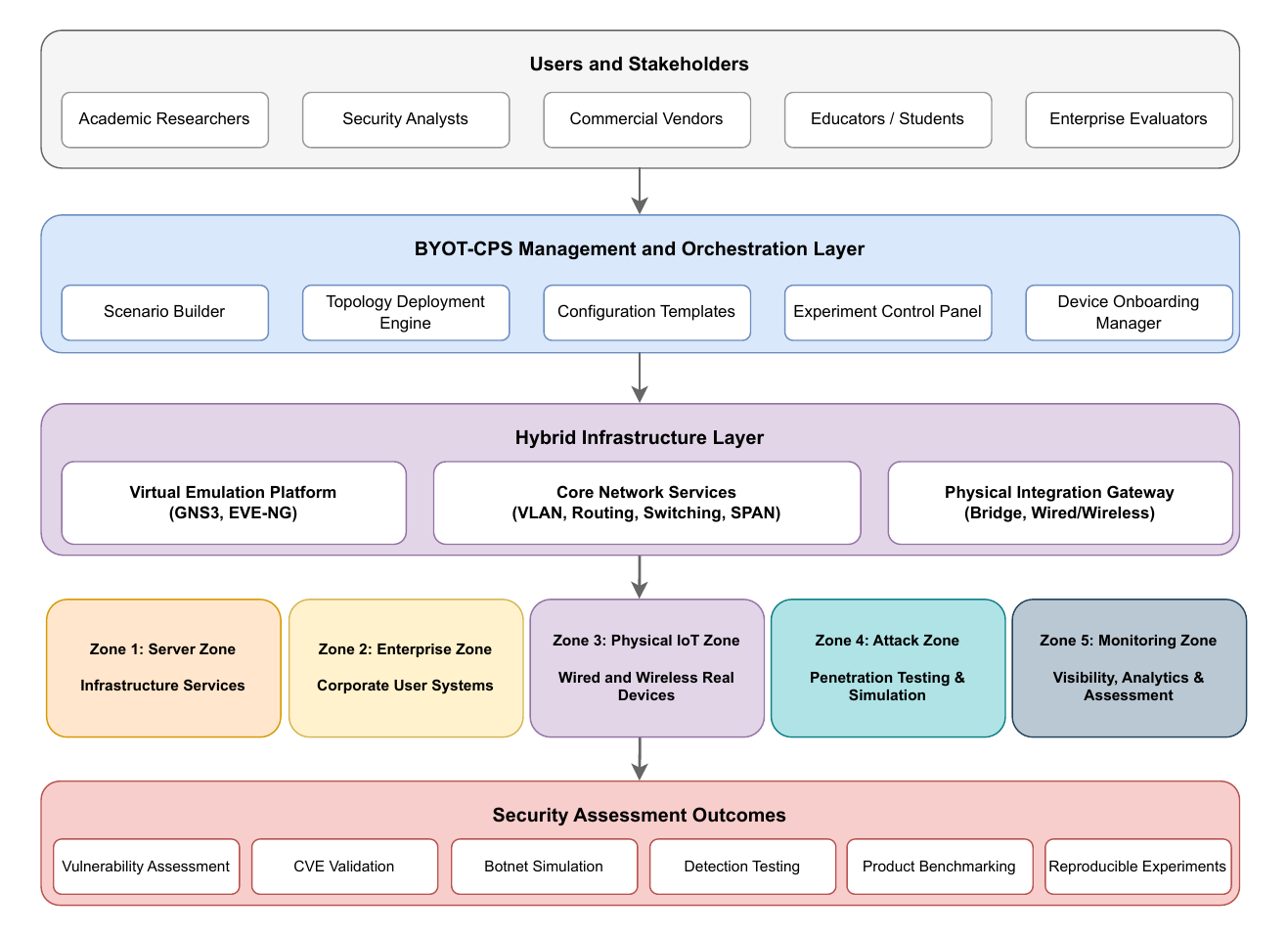}
  \caption{Layered architecture of the BYOT-CPS framework.}
  \label{fig:architecture}
\end{figure*}
Figure~\ref{fig:architecture} presents the layered architecture of BYOT-CPS.
At the top layer are the stakeholder groups: researchers, educators, analysts, vendors, and organisations evaluating candidate security platforms.
Below this is the orchestration layer, responsible for scenario definition, topology deployment, template reuse, and device onboarding.
The hybrid infrastructure layer then combines emulation platforms, routing and switching services, segmentation mechanisms, and physical integration paths.
Beneath this, operational zones host the specific assets required for experimentation. 
The final layer comprises experimental results, including vulnerability validation, attack observation, traffic capture, and platform benchmarking.
\subsection{Reference Topology}
\begin{figure}[!t]
  \centering
  \includegraphics[width=0.98\columnwidth]{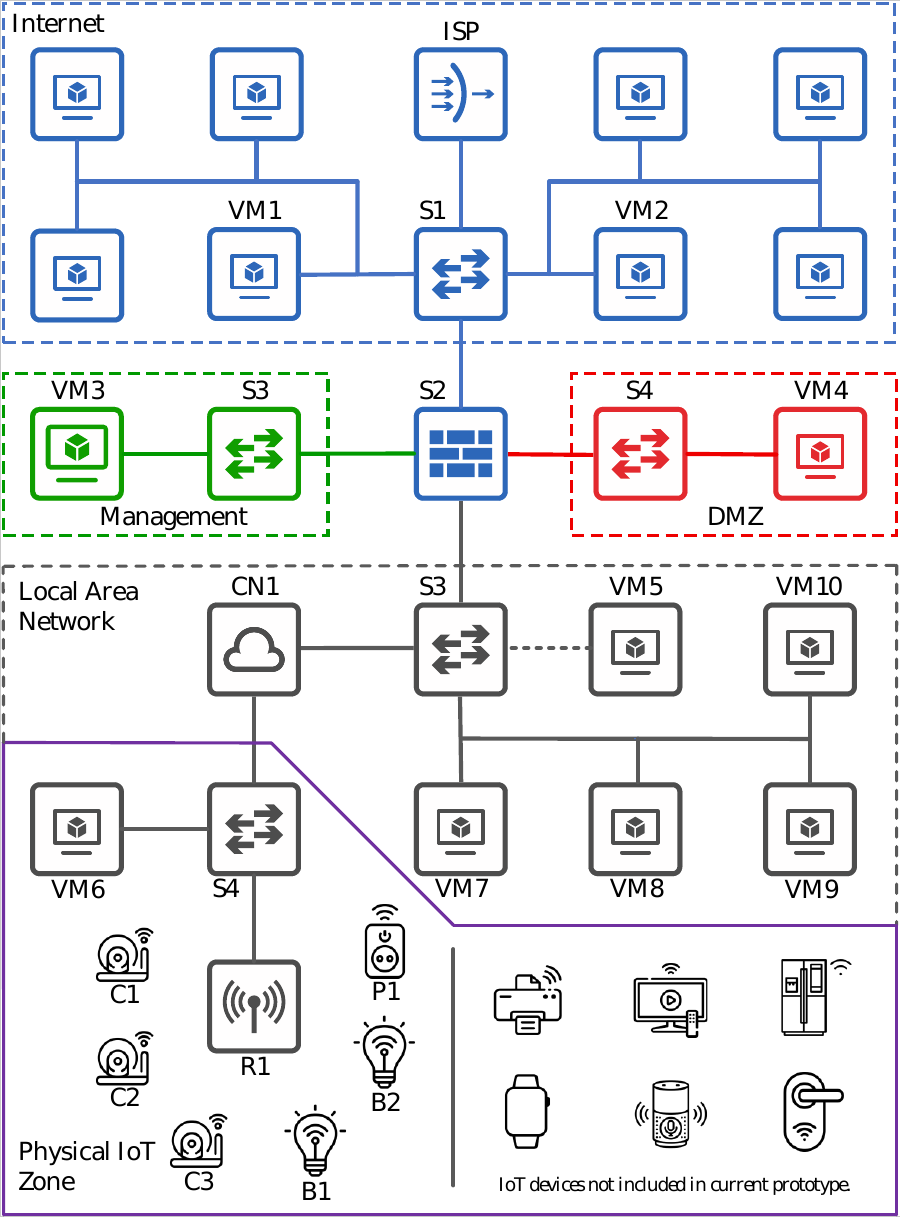}
  \caption{Reference topology of the BYOT-CPS testbed.}
  \label{fig:topology}
\end{figure}

Figure~\ref{fig:topology} shows a representative BYOT-CPS deployment.
The reference topology includes an upstream Internet boundary, pfSense-based firewall and routing, a DMZ hosting server-side services, a management segment for IT administration, an enterprise local area network (LAN), and switching infrastructure supporting the hybrid environment.
A key design decision is the use of a GNS3 Cloud node, an external network switch, and a bridge router to place physical and virtual devices within a common logical addressing scheme while avoiding double network address translation (NAT).
This simplifies management, reduces routing asymmetries, and allows physical IoT devices to interact with virtual servers, attack hosts, and monitoring systems as peer nodes within the same experimental environment.
\subsection{Operational Zones}
The topology is partitioned into five functional zones.
\emph{Server Zone} hosts services such as DNS, web applications, databases, update repositories, and optional command-and-control infrastructure, providing the dependencies required for realistic device and enterprise behaviour.
\emph{Enterprise Zone} represents traditional organisational infrastructure, including Windows and Linux hosts, management consoles, and business-facing services, enabling experiments involving internal reconnaissance, lateral movement, and IT/IoT interaction.
\emph{Physical IoT Devices Zone} contains the real devices integrated into the testbed, including smart bulbs, smart plugs, switches, and IP cameras, and most directly distinguishes BYOT-CPS from emulation-only frameworks.
\emph{Attack Zone} hosts offensive tooling such as Kali Linux, Metasploit, Nmap, credential testing tools, and traffic generation utilities, supporting contained offensive experimentation against both virtual and physical targets.
\emph{Monitoring and Evaluation Zone} includes packet capture sensors, intrusion detection systems, dashboards, and commercial platforms under evaluation; traffic is delivered through a switch port analyzer (SPAN) interface, enabling passive observation without altering the test traffic path.

\subsection{Platform Portability}
Although the prototype reported in this paper was implemented in GNS3, the logical structure of BYOT-CPS was designed with portability to EVE-NG in mind, though this portability has not been empirically validated across platforms.
This portability goal is significant because it may reduce dependence on any single emulation ecosystem and broaden the framework’s applicability across academic, training, and enterprise contexts.

\section{Prototype Validation}\label{sec:evaluation}
This section presents an evaluation of the BYOT-CPS framework across a representative set of IoT security research scenarios.
As this paper constitutes the primary framework publication, the experiments are oriented toward proof-of-concept demonstration of core operational capabilities, including physical device integration, vulnerability assessment, offensive simulation, network traffic analysis, and controlled device manipulation.
Large-scale benchmarking, commercial vendor platform evaluation, and comparative assessment against alternative testbed architectures are deferred to future work.
The prototype evaluation addresses four questions: (1) whether real IoT devices can be integrated into the emulated topology while preserving expected communication behaviour; (2) whether standard offensive security workflows can be executed against both virtual and physical assets; (3) whether the monitoring architecture preserves useful visibility across mixed traffic; and (4) whether the environment can demonstrate observable cyber-physical effects under controlled conditions.
\subsection{Prototype Deployment and Experimental Setup}
The BYOT-CPS prototype was deployed using GNS3 as the primary network emulation platform, chosen for its maturity, widespread adoption within the academic and professional security community, and native support for hybrid virtual-physical integration.
The platform-independent logical design described in Section~\ref{sec:architecture} is intended to enable future reproduction in EVE-NG with minimal reconfiguration, although that portability has not yet been empirically validated.
\begin{table}[!t]
  \begin{threeparttable}
  \caption{Prototype hardware and IoT device inventory used in the BYOT-CPS evaluation.}
  \label{tab:byot-devices}
  \centering  
  \begin{tabular}{>
  {\raggedright\arraybackslash}p{1.75cm}|>
  {\raggedright\arraybackslash}p{2.75cm}|>
  {\raggedright\arraybackslash}p{2.75cm}
  }
  \toprule
  \textbf{Manufacturer} & \textbf{Device Model} & \textbf{Device Type}\\
  \midrule
  Beelink & SER8 AMD Ryzen 7 8745HS, 32GB RAM, 1TB SSD\textsuperscript{\dag} & Host Mini PC for BYOT-CPS \\
  Google  & Nest Wi-Fi Router\textsuperscript{\dag}                          & Wireless Access Point \\
  TP-Link & TL-SG608E\textsuperscript{\dag}                                  & Network Switch \\
  TP-Link & Tapo C210\textsuperscript{\ddag}                                 & IP Camera \\
  TP-Link & Tapo C200\textsuperscript{\ddag}                                 & IP Camera \\
  D-Link  & DCS-930L\textsuperscript{\ddag}                                  & IP Camera (Legacy) \\
  TP-Link & Tapo L530E\textsuperscript{\ddag}                                & Smart Light Bulb \\
  TP-Link & Kasa HS110\textsuperscript{\ddag}                                & Smart Wi-Fi Plug \\
  \bottomrule
  \end{tabular}
  \begin{tablenotes}[para]
      \textsuperscript{\dag}Hardware used to realise BYOT-CPS prototype. 
      \textsuperscript{\ddag} Devices deployed in Physical IoT Zone.
  \end{tablenotes}
  \end{threeparttable}
\end{table}
Table~\ref{tab:byot-devices} summarises the hardware and devices used.
The prototype was provisioned on a Beelink Mini PC.
The TP-Link network switch provides wired network connectivity, while the Google Nest Wi-Fi router, configured as an access point, supports wireless IoT devices.
This configuration was selected for experimental convenience and does not represent a minimum deployment requirement.
A resource-constrained laboratory environment can be realised with approximately 16 GB RAM, 500 GB of storage, a mid-range GPU, and a standard multi-core processor, reflecting the BYOT-CPS design goal of broad institutional deployability.
The physical IoT devices comprised two smart light bulbs, a smart plug, and three IP cameras.
\subsection{Hybrid Connectivity Validation}
The first validation objective was to confirm that COTS IoT devices could be successfully integrated into the emulated network topology while preserving normal communication with virtualised infrastructure components.
The GNS3 Cloud node was used to extend the LAN to an additional USB Ethernet network adapter on the host, which was connected to the network switch.
The Nest Wi-Fi router was configured as a bridge access point, ensuring that wirelessly connected IoT devices remained within the LAN address range.
Both wired and wireless IoT devices were confirmed to be reachable from and able to communicate across the network.
For example, the smart bulb and smart plug could be controlled using the Tapo application, with app-initiated traffic following the configured testbed path to the corresponding devices.
Successful integration of IP cameras, smart bulbs, and the smart plug indicates that BYOT-CPS can bridge the traditional divide between network emulation and physical IoT devices.
Throughout integration testing, physical devices remained discoverable by network scanners, visible to monitoring tools, and accessible through their respective management interfaces.
This suggests that authentic firmware behaviour, including device-initiated cloud communications and vendor-specific protocol activity, was preserved within the testbed environment.

The TP-Link TL-SG608E (\texttt{S4}) provides port-mirroring capability, allowing all wired network traffic to be observed via a designated port.
Because the uplink of the Google Nest Wi-Fi was connected to one of the switch ports, traffic from wirelessly connected IoT devices could also be monitored.
Access to the mirrored port can be provided to commercial platform vendors to demonstrate monitoring and detection capabilities.
Alternatively, switch \texttt{S3} can be implemented as a virtual switch with mirroring capability, such as Cisco IOS on Unix Layer 2; in this configuration, platform vendors can deploy their solutions as virtual appliances within GNS3.

These results demonstrate that BYOT-CPS can integrate real COTS hardware into an emulated topology while preserving behaviours that are difficult to reproduce in simulated, containerised, or purpose-built emulated environments~\cite{gotham-22},~\cite{chernyshev-18}.
\subsection{Vulnerability Assessment and Penetration Testing}
%
%
A primary design objective of BYOT-CPS is to support structured offensive security experimentation within a controlled and isolated environment.
To demonstrate this capability, a series of penetration-testing exercises was conducted using a standard professional toolset: Nmap for host discovery, port enumeration, and OS fingerprinting; Nessus/OpenVAS for automated vulnerability scanning; Metasploit for exploit validation and post-exploitation; and Hydra for credential auditing.
The experiments followed four phases aligned with a typical penetration testing methodology.

\emph{Reconnaissance and Enumeration:} Both virtual and physical devices were successfully identified through active scanning.
Open ports, service versions, OS fingerprints, and management interfaces were enumerated, confirming realistic responses consistent with production environments. 
\emph{Vulnerability Identification:} Automated tools revealed common weaknesses across IoT devices, including outdated firmware, exposed services, and insecure default configurations.
These findings align with established IoT vulnerability profiles~\cite{sicari-15,wei-25}, thereby supporting the realism of the testbed.
\emph{Credential and Access Testing:} Controlled authentication testing demonstrated the risks posed by default credentials, a vulnerability class central to large-scale IoT compromise campaigns~\cite{mirai-17,kolias-17}.
\emph{Controlled Exploitation:} Selected proof-of-concept exploits were executed using Metasploit and manual techniques to validate practical exploitability.
For example, controlled exploitation of the Tapo C200 and legacy D-Link DCS-930L enabled observable behaviours such as service activation and device manipulation as detailed in Table~\ref{tab:byot-cps-vulns}.

These results demonstrate that BYOT-CPS supports representative penetration-testing workflows across heterogeneous target environments comprising both virtualised systems and real IoT hardware, reinforcing its utility for research, professional training, and defensive validation.
\begin{table}[!t]
  \begin{threeparttable}
  \caption{Vulnerabilities of IoT Devices}
  \label{tab:byot-cps-vulns}
  \centering
  \footnotesize
  \setlength{\tabcolsep}{3pt}
  \begin{tabular}{>
  {\raggedright\arraybackslash}p{1.45cm}|>
  {\raggedright\arraybackslash}p{3.55cm}|>
  {\raggedright\arraybackslash}p{2.10cm}
  }
  \toprule
  \textbf{Device} & \textbf{Vulnerability} & \textbf{CVE}\\
  \midrule
  Tapo C210  & Password recovery authentication bypass & CVE-2023-35717 \\
  Tapo C210  & ActiveCells stack-based buffer overflow & CVE-2023-41184 \\
  Tapo C210  & Oversized URL in HTTP parser            & CVE-2026-0919 \\
  Tapo C210  & Stack-based overflow                    & CVE-2026-34122 \\
  Tapo C200  & Unauthenticated remote code execution   & CVE-2021-4045 \\
  Tapo C200  & Hardcoded encryption keys               & CVE-2023-27126 \\
  Tapo C200  & ONVIF SOAP parser buffer overflow       & CVE-2025-8065\textsuperscript{\dag} \\
  Tapo C200  & Integer overflow DoS                    & CVE-2025-14299\textsuperscript{\dag} \\
  Tapo C200  & connectAP authentication bypass         & CVE-2025-14300\textsuperscript{\dag} \\
  Tapo C200  & Password hash leak                      & CVE-2025-14553 \\
  DCS-930L   & OS command injection vulnerability      & CVE-2016-11021\textsuperscript{\dag} \\
  \bottomrule
  \end{tabular}
  \begin{tablenotes}[para]
    \textsuperscript{\dag} Vulnerability validated in the BYOT-CPS prototype environment. Other entries denote vulnerabilities referenced from public sources and included to contextualise the device classes considered; they were not validated experimentally in this study.
  \end{tablenotes}
  \end{threeparttable}
\end{table}
\subsection{Mirai-Style Denial-of-Service Attack Scenario}
\begin{figure}
  \centering
  \includegraphics[width=0.85\columnwidth]{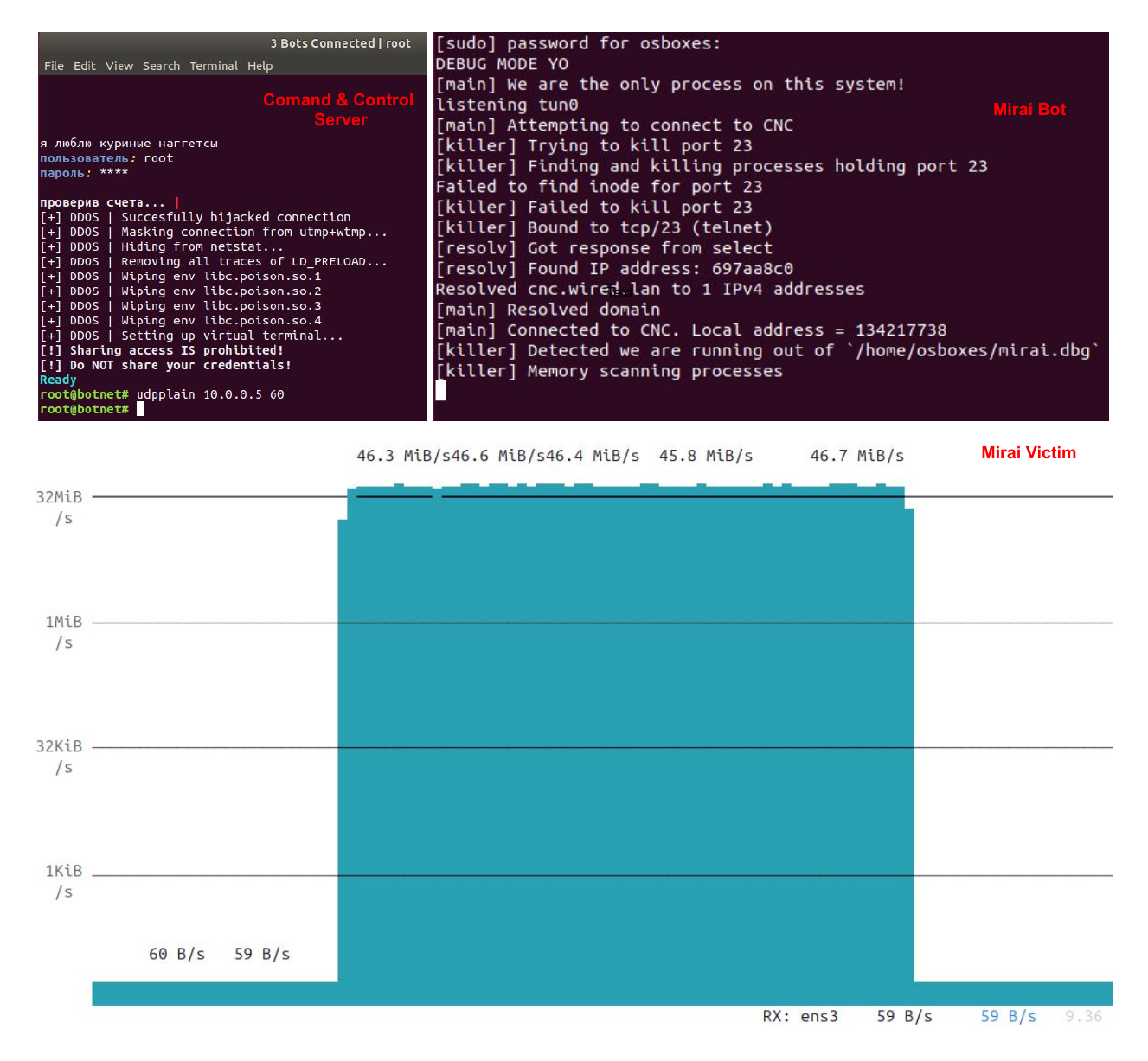}
  \caption{Mirai-style denial-of-service scenario in BYOT-CPS}
  \label{fig:mirai}
\end{figure}

%
%
To evaluate BYOT-CPS under a coordinated attack scenario, a Mirai-style denial-of-service experiment was constructed within the testbed.
The objective was not to cause destructive disruption, but to provide a controlled and observable demonstration of sustained attack traffic generated by compromised hosts.
In this setup, three LAN hosts (\texttt{VM7}, \texttt{VM8}, \texttt{VM9}) were infected with the Mirai malware and directed by a command-and-control server (\texttt{VM2}) to perform a UDP-based DoS attack against a target host (\texttt{VM10}) for 60 seconds.
As shown in Figure~\ref{fig:mirai}, the attack traffic reached approximately 46 Mbps.
The bots were subsequently redirected to target a smart plug (\texttt{P1}), which then became unresponsive to control commands.

This experiment is useful in teaching and training settings, illustrating how botnet-style malware can generate sustained attack traffic within a contained environment.
The scenario reflects core operational characteristics of the 2016 Mirai botnet and related campaigns~\cite{kolias-17}, allowing students and analysts to observe IoT-driven attack dynamics without external network exposure.
These results demonstrate that BYOT-CPS can safely replicate Mirai-style botnet behaviour and denial-of-service activity within a contained research environment.
\subsection{Traffic Monitoring and Visibility}
A further objective was to evaluate BYOT-CPS as a monitoring and dataset-generation environment.
Traffic capture was conducted using Wireshark, \texttt{tcpdump}, and TShark on the SPAN port, enabling simultaneous inspection of traffic from physical IoT devices and virtual infrastructure components. 
Observed traffic included routine device beaconing, cloud communication, service discovery, management activity, and anomalous scanning or flooding associated with attack exercises.
Access to authentic mixed-environment traffic supports behavioural analysis, intrusion detection evaluation, and future dataset generation using tools such as GothX~\cite{gothx-24} and MQTTset~\cite{mqttset-20}.
It also enables the evaluation of commercial IoT security platforms under passive monitoring conditions, where effective detection depends on both attack traffic and normal device context.
Such contextual richness is difficult to reproduce in purely emulated environments.

\subsection{Comparative Visibility Case Study: Emulation-Only vs. BYOT-CPS}
To illustrate the value of real-device participation, a qualitative case study compared an emulation-only baseline with the BYOT-CPS hybrid deployment.
The objective was to assess whether security-relevant observations in the hybrid environment would be absent, reduced, or distorted in a virtual-only approximation.

Both scenarios used the same network path, monitoring point, and attack host.
In the baseline, the target was a lightweight virtual IoT camera with limited service exposure.
In BYOT-CPS, the targets were real devices integrated via the bridge architecture, including IP cameras, smart bulbs, and a smart plug. 
Traffic was captured during both idle operation and active probing, with additional observation during controlled Mirai activity in the hybrid setup.

The comparison focused on service realism, protocol diversity, background communication, and the visibility of behaviour relevant to security monitoring.
Table~\ref{tab:case-study} summarises the results. 
While both environments supported basic IP-level experimentation, BYOT-CPS exposed richer and more representative behaviour, including authentic management interfaces, device-specific traffic patterns, cloud communication, and observable cyber-physical effects, features absent or simplified in the emulation-only case.

These differences are significant because security evaluation depends on behavioural fidelity rather than simple reachability.
The value of a hybrid testbed lies not only in incorporating physical devices, but in producing mixed traffic traces in which benign activity, management behaviour, and malicious actions coexist in a realistic form.
This comparison is qualitative and intended as an illustrative case study rather than a benchmark.
Future work should provide packet-level evidence, including identified services, outbound communication patterns, vendor-specific traffic, and the observability of botnet-related activity.
\begin{table}[t]
  \caption{Comparative visibility case study between an emulation-only baseline and BYOT-CPS.}
  \label{tab:case-study}
  \centering
  \begin{tabular}{>
  {\raggedright\arraybackslash}p{4.75cm}|>
  {\raggedright\arraybackslash}p{1.25cm}|>
  {\raggedright\arraybackslash}p{1.40cm}
  }
  \toprule
  \textbf{Observation} & \textbf{Emulation} & \textbf{BYOT-CPS} \\
  \midrule
  Network reachability                  & Yes     & Yes \\
  Service fingerprinting                & Partial & Yes \\
  Device-specific traffic visible       & No      & Yes \\
  Background/cloud communications       & No      & Yes \\
  Native management interface behaviour & No      & Yes \\
  Observable cyber-physical effect      & No      & Yes \\
  Mirai traffic realism                 & Limited & Yes \\
  \bottomrule
  \end{tabular}
\end{table}
\begin{table}[!t]
  \caption{Representative controlled device interactions and observable effects in the BYOT-CPS prototype.}
  \label{tab:byot-cps-effects}
  \centering
  \footnotesize
  \setlength{\tabcolsep}{3pt}
  \begin{tabular}{>
  {\raggedright\arraybackslash}p{1.45cm}|>
  {\raggedright\arraybackslash}p{2.95cm}|>
  {\raggedright\arraybackslash}p{2.85cm}
  }
  \toprule
  \textbf{Device} & \textbf{Interaction} & \textbf{Observed effect}\\
  \midrule
  Tapo L530E & Customised control script from \texttt{VM9} & On/off state change \\
  Kasa HS110 & Customised control script from \texttt{VM9} & On/off state change \\
  Tapo C200  & Controlled exploit activity                 & Exploit behaviour demonstrated \\
  DCS-930L   & Controlled exploit activity                 & Telnet service enabled; device subsequently used in Mirai-style scenario \\
  \bottomrule
  \end{tabular}
\end{table}
\subsection{Device Manipulation and Real-World Impact Demonstration}
A notable set of experiments examined the tangible consequences of IoT device compromise beyond conventional data exfiltration.
Within the isolated testbed, authorised manipulation of physical devices was conducted following simulated access acquisition.
Demonstrated actions included remote toggling of smart bulbs, disruption of IP camera operation, and modification of device states via exposed management interfaces as detailed in Table~\ref{tab:byot-cps-effects}.
Although performed under controlled and ethically governed conditions, these scenarios reflect realistic threat contexts.
Equivalent access in operational environments could enable adversaries to disrupt surveillance systems, building management infrastructure, or other IoT-connected services, with impacts extending beyond data confidentiality.
These results highlight that IoT compromise can produce physical and operational consequences distinct from traditional IT incidents, and that evaluation frameworks which ignore this dimension provide only a partial representation of IoT risk.
\subsection{Traffic Interception and Network Visibility}
BYOT-CPS was further evaluated as a network monitoring and traffic analysis environment, assessing its suitability for behavioural research, protocol analysis, and intrusion detection experimentation.
Traffic capture was performed using Wireshark for interactive inspection and \texttt{tcpdump}/TShark for command-line capture and filtering.
Traffic was mirrored to a SPAN port, in line with the passive monitoring architecture described in Section~\ref{sec:architecture}, enabling simultaneous observation of physical IoT and virtual infrastructure traffic.
Observed phenomena included communication patterns across device classes, protocol usage, periodic beaconing and cloud connectivity, instances of cleartext or inadequately protected application-layer traffic, and anomalous scanning or flooding associated with attack scenarios.
The ability to capture authentic traffic from real IoT devices, rather than synthetic or replayed traces, provides clear value for behavioural baselining, anomaly detection, intrusion detection evaluation, and firmware communication reverse engineering.
This capability distinguishes BYOT-CPS from emulation-only environments, where traffic realism is typically limited or approximate.
\subsection{Reproducibility, Flexibility, and Scalability}
The cumulative experimental programme illustrates how BYOT-CPS can be deployed across different resource constraints and research objectives.
At the lower end, a minimal configuration can be deployed by a single researcher using a small set of virtual nodes and limited IoT devices.
At the upper end, the same architecture can scale to multi-zone environments supporting larger device populations, commercial monitoring platforms, and concurrent users.
The software-defined topology enables operational flexibility, including scenario cloning, dynamic addition or removal of zones, hot swapping of virtual or physical devices without topology redesign, and template-based scenario reuse for teaching and repeatable experiments.
These properties demonstrate how BYOT-CPS can support reproducibility, extensibility, and scalability in future larger-scale deployments, rather than establishing them as fully benchmarked outcomes in the current work.
\subsection{Summary of Validation}
Taken together, these experiments indicate that the BYOT-CPS prototype supports the core capabilities expected of a hybrid IoT security testbed, including the integration of virtual infrastructure with real IoT devices, support for realistic offensive workflows against heterogeneous targets, the generation and observation of contained botnet traffic, passive visibility across mixed physical and virtual environments, and the demonstration of controlled cyber-physical effects.
The prototype therefore provides feasibility evidence for the BYOT-CPS design and motivates future work involving larger-scale experimentation, comparative evaluation, and systematic assessment of security platforms.
\section{Discussion, Limitations, and Future Work}\label{sec:discussion}
The current prototype demonstrates the viability of BYOT-CPS, but should be understood as an initial framework implementation rather than a complete end state.
In its present form, BYOT-CPS constitutes a research infrastructure contribution accompanied by an initial feasibility validation, rather than a comprehensive benchmark study.

\subsection{Current Limitations}
Several limitations remain.
First, although the physical device set is heterogeneous, it remains modest in scale and does not yet cover the breadth of enterprise, industrial, building automation, and healthcare-related IoT systems required for comprehensive benchmarking.
Second, while the architecture was designed with EVE-NG portability in mind, empirical cross-platform validation remains future work.
Third, the commercial platform evaluation use case has been defined and technically supported, but systematic benchmarking of platforms such as Forescout, Claroty, and Armis is beyond the scope of this work.
Finally, scenario orchestration remains partly manual, introducing overhead for larger or repeated experiments.
\subsection{Threats to Validity}
Several threats to validity apply. 
First, although the device set is diverse, it remains limited in scale and does not fully represent industrial, medical, or building-management IoT deployments. 
Second, the evaluation is feasibility-oriented, demonstrating representative security workflows without large-scale benchmarking against alternative approaches.
Third, claims regarding portability and vendor-neutral evaluation are supported by architectural design and workflow validation rather than cross-site studies or systematic product comparisons. 
Finally, all experiments were conducted in a controlled research setting, and operational constraints may differ in larger deployments.
These limitations constrain the conclusions that can be drawn; this work presents a research infrastructure contribution with initial empirical validation rather than a comprehensive comparative assessment.
\subsection{Future Work}
%
Several future directions emerge from this work.
First, conducting repeatable comparative studies across emulation-only, physical-only, and hybrid environments using common attack scenarios and monitoring pipelines would provide deeper insight into system behaviour.
Second, the physical device set can be expanded to incorporate additional protocols and device classes, including ZigBee, Z-Wave, industrial sensors, and other operational technology endpoints.
Third, platform portability should be formally validated across both GNS3 and EVE-NG environments.
Fourth, integrating traffic generation frameworks such as GothX~\cite{gothx-24} and MQTTset~\cite{mqttset-20} would support the creation of labelled datasets for further analysis~\cite{dataset-19}.
Fifth, orchestration tooling could be developed to automate topology deployment, device onboarding, and the execution of multi-stage attack scenarios.
Finally, BYOT-CPS offers a strong foundation for the vendor-neutral evaluation of commercial IoT security solutions, enabling systematic assessment of asset discovery, protocol identification, and behavioural analysis under controlled and reproducible conditions.
\section{Conclusion}\label{sec:conclusion}
This paper presented Build Your Own Cyber-Physical Systems Testbed (BYOT-CPS), a hybrid framework designed to bridge the gap between scalable virtual experimentation and realistic IoT device behaviour.
BYOT-CPS integrates COTS IoT hardware with virtualised network infrastructure, enabling physical devices to be studied, attacked, monitored, and evaluated within a controlled, reproducible environment.
Six requirements were defined for hybrid IoT security testbeds: fidelity, heterogeneity, scalability, reproducibility, extensibility, and independence. Existing approaches typically address only a subset of these.

BYOT-CPS addresses this gap by combining real-device fidelity with flexible emulated topologies and by establishing vendor-neutral platform evaluation as a primary design objective.
Prototype evaluation showed the ability to integrate real devices into a shared topology, support representative offensive workflows, instantiate contained Mirai-style denial-of-service activity, provide passive monitoring visibility, and capture cyber-physical effects. 
These results provide evidence of feasibility and practical utility rather than a comprehensive benchmark.

As IoT deployments continue to expand, the ability to study real device behaviour without sacrificing experimental control is increasingly important.
BYOT-CPS contributes an extensible foundation for this purpose, with value for researchers, educators, defenders, and organisations seeking an independent basis for IoT security platform evaluation.
\bibliographystyle{IEEEtran}
\bibliography{references}

\begin{thebibliography}{10}
\providecommand{\url}[1]{#1}
\csname url@samestyle\endcsname
\providecommand{\newblock}{\relax}
\providecommand{\bibinfo}[2]{#2}
\providecommand{\BIBentrySTDinterwordspacing}{\spaceskip=0pt\relax}
\providecommand{\BIBentryALTinterwordstretchfactor}{4}
\providecommand{\BIBentryALTinterwordspacing}{\spaceskip=\fontdimen2\font plus
\BIBentryALTinterwordstretchfactor\fontdimen3\font minus \fontdimen4\font\relax}
\providecommand{\BIBforeignlanguage}[2]{{%
\expandafter\ifx\csname l@#1\endcsname\relax
\typeout{** WARNING: IEEEtran.bst: No hyphenation pattern has been}%
\typeout{** loaded for the language `#1'. Using the pattern for}%
\typeout{** the default language instead.}%
\else
\language=\csname l@#1\endcsname
\fi
#2}}
\providecommand{\BIBdecl}{\relax}
\BIBdecl

\bibitem{iot-analytics-25}
\BIBentryALTinterwordspacing
S.~Sinha. State of {IoT} 2025: Number of connected {IoT} devices growing 14\% to 21.1 billion globally. [Online]. Available: \url{https://iot-analytics.com/number-connected-iot-devices/}
\BIBentrySTDinterwordspacing

\bibitem{statista-25}
Statista, ``{IoT} connected devices worldwide 2025--2034,'' \url{https://www.statista.com/statistics/1183457/iot-connected-devices-worldwide/}, 2026.

\bibitem{mirai-17}
M.~Antonakakis, T.~April, M.~Bailey, M.~Bernhard, E.~Bursztein, J.~Cochran, Z.~Durumeric, J.~A. Halderman, L.~Invernizzi, M.~Kallitsis, D.~Kumar, C.~Lever, Z.~Ma, J.~Mason, D.~Menscher, C.~Seaman, N.~Sullivan, K.~Thomas, and Y.~Zhou, ``Understanding the mirai botnet,'' in \emph{Proceedings of the 26th USENIX Conference on Security Symposium}, ser. SEC'17.\hskip 1em plus 0.5em minus 0.4em\relax USA: USENIX Association, 2017, p. 1093–1110.

\bibitem{meneghello-19}
F.~Meneghello, M.~Calore, D.~Zucchetto, M.~Polese, and A.~Zanella, ``Iot: Internet of threats? a survey of practical security vulnerabilities in real iot devices,'' \emph{IEEE Internet of Things Journal}, vol.~6, no.~5, pp. 8182--8201, 2019.

\bibitem{cooja-06}
F.~Osterlind, A.~Dunkels, J.~Eriksson, N.~Finne, and T.~Voigt, ``Cross-level sensor network simulation with cooja,'' in \emph{Proceedings. 2006 31st IEEE Conference on Local Computer Networks}, 2006, pp. 641--648.

\bibitem{ns3-10}
G.~F. Riley and T.~R. Henderson, \emph{The ns-3 Network Simulator}.\hskip 1em plus 0.5em minus 0.4em\relax Berlin, Heidelberg: Springer Berlin Heidelberg, 2010, pp. 15--34.

\bibitem{omnet-08}
A.~Varga and R.~Hornig, ``An overview of the {OMNeT++} simulation environment,'' ser. Simutools '08, Brussels, BEL, 2008.

\bibitem{core-08}
J.~Ahrenholz, C.~Danilov, T.~R. Henderson, and J.~H. Kim, ``{CORE}: A real-time network emulator,'' in \emph{MILCOM 2008 - 2008 IEEE Military Communications Conference}, 2008, pp. 1--7.

\bibitem{mininet-10}
B.~Lantz, B.~Heller, and N.~McKeown, ``A network in a laptop: rapid prototyping for software-defined networks,'' in \emph{Proceedings of the 9th ACM SIGCOMM Workshop on Hot Topics in Networks}, ser. Hotnets-IX.\hskip 1em plus 0.5em minus 0.4em\relax New York, NY, USA: Association for Computing Machinery, 2010.

\bibitem{gotham-22}
X.~Saez-de Camara, J.~L. Flores, C.~Arellano, A.~Urbieta, and U.~Zurutuza, ``{Gotham Testbed: A Reproducible IoT Testbed for Security Experiments and Dataset Generation},'' \emph{IEEE Transactions on Dependable and Secure Computing}, vol.~21, no.~01, pp. 186--203, Jan. 2024.

\bibitem{epic-13}
C.~Siaterlis, B.~Genge, and M.~Hohenadel, ``{EPIC}: A testbed for scientifically rigorous cyber-physical security experimentation,'' \emph{IEEE Transactions on Emerging Topics in Computing}, vol.~1, no.~2, pp. 319--330, 2013.

\bibitem{attack-graphs-06}
R.~Lippmann, K.~Ingols, C.~Scott, K.~Piwowarski, K.~Kratkiewicz, M.~Artz, and R.~Cunningham, ``Validating and restoring defense in depth using attack graphs,'' in \emph{MILCOM 2006 - 2006 IEEE Military Communications conference}, 2006, pp. 1--10.

\bibitem{islam-15}
S.~M.~R. Islam, D.~Kwak, M.~H. Kabir, M.~Hossain, and K.-S. Kwak, ``The internet of things for health care: A comprehensive survey,'' \emph{IEEE Access}, vol.~3, pp. 678--708, 2015.

\bibitem{gns3}
{GNS3 Technologies}, ``Gns3 documentation,'' \url{https://docs.gns3.com/}, 2026.

\bibitem{eve-ng-25}
{EVE-NG Ltd}, ``{EVE-NG}: The emulated virtual environment for network, security and devops professionals,'' \url{https://www.eve-ng.net/}, 2026.

\bibitem{hybrid-23}
K.~E. Balto, M.~M. Yamin, A.~Shalaginov, and B.~Katt, ``Hybrid iot cyber range,'' \emph{Sensors}, vol.~23, no.~6, 2023.

\bibitem{container-19}
M.~M.~H. Onik, C.-S. Yang, M.~A. Razzaque, and M.~A. Serhani, ``Container-based intrusion detection systems for the internet of things,'' \emph{Sensors}, vol.~19, no.~23, p. 5277, 2019.

\bibitem{mqttset-20}
I.~Vaccari, G.~Chiola, M.~Aiello, M.~Mongelli, and E.~Cambiaso, ``Mqttset, a new dataset for machine learning techniques on mqtt,'' \emph{Sensors}, vol.~20, no.~22, 2020.

\bibitem{gothx-24}
M.~Poisson, R.~Carnier, and K.~Fukuda, ``Gothx: a generator of customizable, legitimate and malicious iot network traffic,'' in \emph{Proceedings of the 17th Cyber Security Experimentation and Test Workshop}, ser. CSET '24, New York, NY, USA, 2024, p. 65–73.

\bibitem{emulytics}
{Sandia National Laboratories}, ``{Emulytics} — {Cyber} at {Sandia},'' \url{https://www.sandia.gov/emulytics/}, 2026.

\bibitem{chernyshev-18}
M.~Chernyshev, Z.~Baig, O.~Bello, and S.~Zeadally, ``Internet of things (iot): Research, simulators, and testbeds,'' \emph{IEEE Internet of Things Journal}, vol.~5, no.~3, pp. 1637--1647, 2018.

\bibitem{sicari-15}
S.~Sicari, A.~Rizzardi, L.~Grieco, and A.~Coen-Porisini, ``Security, privacy and trust in internet of things: The road ahead,'' \emph{Computer Networks}, vol.~76, pp. 146--164, 2015.

\bibitem{wei-25}
Z.~Wei, Q.~Wei, Y.~Geng, and Y.~Yang, ``A survey on iot security: Vulnerability detection and protection,'' in \emph{Proceedings of the 2024 International Conference on Artificial Intelligence of Things and Computing}, ser. AITC '24.\hskip 1em plus 0.5em minus 0.4em\relax New York, NY, USA: Association for Computing Machinery, 2025, p. 1–8.

\bibitem{kolias-17}
C.~Kolias, G.~Kambourakis, A.~Stavrou, and J.~Voas, ``{DDoS in the IoT: Mirai and Other Botnets},'' \emph{Computer}, vol.~50, no.~7, pp. 80--84, 2017.

\bibitem{dataset-19}
N.~Koroniotis, N.~Moustafa, E.~Sitnikova, and B.~Turnbull, ``Towards the development of realistic botnet dataset in the internet of things for network forensic analytics: Bot-{IoT} dataset,'' \emph{Future Generation Computer Systems}, vol. 100, pp. 779--796, 2019.

\end{thebibliography}
\end{document}